\documentclass[10pt,twocolumn]{article}
\AtBeginDocument{%
  \providecommand\BibTeX{{%
    \normalfont B\kern-0.5em{\scshape i\kern-0.25em b}\kern-0.8em\TeX}}}

\usepackage{multirow}
\usepackage{comment}
\usepackage{caption}
\usepackage{subcaption}
\usepackage{xspace}
\usepackage[most]{tcolorbox}
\PassOptionsToPackage{hyphens}{url}\usepackage{hyperref}

\newcommand{\one}{({\em i}\/)\xspace}
\newcommand{\two}{({\em ii}\/)\xspace}
\newcommand{\three}{({\em iii}\/)\xspace}
\newcommand{\four}{({\em iv}\/)\xspace}

\def\eg{\emph{e.g.}\xspace}
\def\ie{\emph{i.e.}\xspace}

\def\vs{\emph{vs.}\xspace}

\newcommand{\pb}[1]{\vspace{0.75ex}\noindent{\bf \em #1}\hspace*{.3em}}

\graphicspath{ {./images/} }

\begin{document}

\title{An Analysis of Twitter Discourse on the War Between Russia and Ukraine}

\author{Haris Bin Zia$^1$, Ehsan Ul Haq$^2$, Ignacio Castro$^{1}$, Pan Hui$^3$, and Gareth Tyson$^{1,3}$\\[0.5ex]
\normalsize
$^1$Queen Mary University of London, $^2$Hong Kong University of Science and Technology, \\
\normalsize
$^3$Hong Kong University of Science and Technology (GZ)\\[0.5ex]
\small
Corresponding author: h.b.zia@qmul.ac.uk
}
\maketitle

\begin{abstract}
On the 21st of February 2022, Russia recognised the Donetsk People’s Republic and the Luhansk People’s Republic, three days before launching an invasion of Ukraine. Since then, an active debate has taken place on social media, mixing organic discussions with coordinated information campaigns. The scale of this discourse, alongside the role that information warfare has played in the invasion, make it vital to better understand this ecosystem.
We therefore present a study of pro-Ukrainian \vs pro-Russian discourse through the lens of Twitter. We do so from two perspectives: \one~the content that is shared; and \two~the users who participate in the sharing. We first explore the scale and nature of conversations, including analysis of hashtags, toxicity and media sharing. We then study the users who drive this, highlighting a significant presence of new users and bots.
\end{abstract}

\section{Introduction}

Tensions between Russia and Ukraine have been rising since the 2014 annexation of the Crimean Peninsula~\cite{wikipediacrimes}. On 21st February 2022, the conflict escalated after Russia recognised the Donetsk People’s Republic and the Luhansk People’s Republic (two Ukrainian breakaway statelets as independent regions). Later on 24th February, the Russian government began a "special military operation" in Ukraine, amounting to a full-scale invasion~\cite{wikipediukraine}. 
Anecdotally, this has been accompanied by a flurry of social media activity both by citizens and coordinated state actors. 

This is not a new phenomena. 
Increasingly, social networks have played an important role in conflicts. 
For example, citizens might use social media platforms to share their opinions, call for assistance, and report human rights violations. 
Whereas, similarly, state actors might misuse such platforms to spread propaganda and misinformation~\cite{haq2022weapon}.
This has serious consequences, and undermines faith in public discourse~\cite{zannettou2019disinformation}.

Considering the severity of the Russia-Ukraine crisis, we argue it is vital to gain a better understanding of the ongoing discourse. Key questions include 
\one~what type of discussions occur on Twitter and how much are they influenced by real-world events? 
\two~How much, if at all, are conversations around the conflict becoming toxic? 
\three~Do countries withhold content in their territories during the time of conflict, and does this have political goals? 
and 
\four~How, if at all, do inauthentic and automated accounts amplify certain narratives during conflict?

To explore these questions, we perform the first large-scale analysis of political discourse on Twitter during the Russia-Ukraine war. 
We focus on Twitter due to the active role it plays in information warfare~\cite{haq2022weapon}.
Our analysis is performed across two major axes.
First, we explore the types of content shared that discuss the war. (\S\ref{sec:contentperspective}), before inspecting the users who share such material (\S\ref{sec:userperspetive}).

We find that pro-Ukrainian content is more prevalent on Twitter than pro-Russian content, and that the online discourse is tightly coupled with the real-world political events (\S\ref{sec:hashtags}).
For instance, we see significant spikes in posting activity around the start of invasion and the UN General Assembly vote. 
We also observe differences across regions and languages (\S\ref{sec:language}).
Unsurprisingly, Western languages dominate the dataset, with a large volume of posts from Europe and the US. 
In-line with common thinking, our results show that Ukraine gathers support from the global West, whilst users in the sub-continent and Africa are more inclined towards the Russian narrative.
Posting of URLs is commonplace (largely news media), but interestingly we find regular hijacking of the narrative by people posting links to things like pornography and spam (\S\ref{sec:urls}).

We also inspect the reactions of the two countries to the opposing narratives. 
We observe evidence of heightened toxicity within the tweets (\S\ref{sec:toxicity}), particularly for pro-Russian content. 
14.7\% of the tweets containing pro-Russian hashtags are toxic compared to just 5.9\% of pro-Ukrainian tweets.
However, a more granular view shows that tweets containing the word ``Putin`` are more toxic than that of any global leaders. Often this involves pro-Ukranian users 'hijacking' pro-Russian hashtags as part of their critique.
We conjecture that this toxicity may result in countries using Twitter's withheld feature (which blocks particular users and posts from appearing on a country's timelines).
Indeed, we find that tweets are withheld from both sides as a consequence of the conflict (\S\ref{sec:witheld}).
Countries in Europe withhold the most tweets, and the largest fraction of these originate from Russia. 
For example, we see common blocking of posts from state-owned media \eg @RTarabic and @RT\_russian.
We also discover that a significant surge of posts come from new users (\S\ref{sec:newusers}), which makes us suspect bots may play an active role in driving discourse (\S\ref{sec:bots}).
In-line with anecdotal evidence, we find that a large number of pro-Russian tweets are driven by bots. Surprisingly, we also find highly automated pro-Ukrainian bots too. We conclude by inspecting the retweet graphs of users who are active in the debate (\S\ref{sec:interactiongraph}), highlighting that accounts that use pro-Ukrainian hashtags have clustered more tightly compared to pro-Russian hashtags.

\section{Data Collection}

\begin{figure*}
     \centering
     \subfloat[][]{
     \includegraphics[width=.5\linewidth, height=5cm]{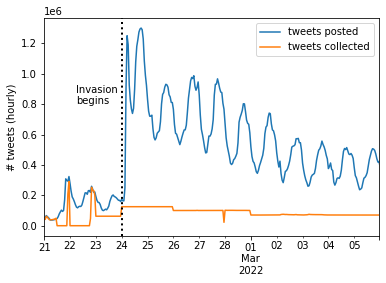} \label{lab:tweetspostedcollected}}
     \subfloat[][]{\includegraphics[width=.5\linewidth, height=5cm]{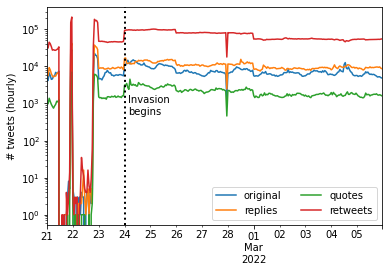}\label{lab:tweettypes}}
     \caption{(a) Number of tweets collected vs. number of tweets posted in each hour. (b) Number of tweets (per hour) of different types. Note log scale on y-axis.}
     \label{steady_state_1}
\end{figure*}

\pb{Tweet Collection.}
We use Twitter’s Search API~\cite{twittersearchapi} to collect tweets related to the conflict from February 21, 2022 to March 05, 2022.
We do not continue data collection beyond this point because the Russian state blocked Twitter in March.
We leverage a list of keywords bootstrapped from trending words \& hashtags on Twitter and recent studies around Russian-Ukraine war~\cite{pierri2022propaganda,pierri2022does} to search tweets.
These keywords are `ukraine', `russia', `putin', `zelensky', `kyiv', `russian', `ukrainian', `keiv' and translation of these keywords in Ukrainian \& Russian language. 

\pb{Location Tags.}
Twitter allows its users to tag their tweets with a location 
This contains the geo-coordinates of the location from where the user posted the tweet. In our dataset, $<1$\% of the tweets contain this information. 
Thus, following \cite{jiang2020political,chen2022tweets}, we rely on the location manually specified by users in their profiles. User specified locations are free-form strings that we map to a country code using OpenStreetMap Search API~\cite{openstreetmapsearchapi}.
Upon applying this methodology, we tag 84.43\% of tweets with their location.

\pb{Toxicity.}
To analyze the extent of toxic material in our dataset, we label all tweets containing pro-Russian and pro-Ukrainian hashtags using Google Jigsaw’s Perspective API~\cite{perspectiveapi}. 
For a given tweet, Perspective returns a score between 0 and 1 for its toxicity (0 $=$ non-toxic). Specifically, we use the API's ``SEVERE\_TOXICITY`` attribute that defines toxicity as ``a very hateful, aggressive, disrespectful comment or otherwise very likely to make a user leave a discussion or give up on sharing their perspective.``

\pb{Data Summary.}
Figure~\ref{lab:tweetspostedcollected} presents the number of tweets collected and the number of tweets posted per hour. 
The percentage of collected tweets on different days varies between 12.8-39.9\% of the tweets posted. In total, our dataset consists of 25,177,195 tweets posted by 5,520,235 users. Not surprisingly, the majority of the tweets in our dataset are retweets followed by replies, original and quotes. 
For context, the temporal distribution of different tweet types is shown in Figure~\ref{lab:tweettypes}. 

\pb{Ethical Considerations.} The data in this work has been collected only using Twitter API respecting the Twitter's terms of service. We have exclusively collected the publicly available data, followed well established ethical procedures for social data, and obtained an approval from the ethics committee at the first author’s institution. We have not identified individual users with the exception of a handful of public accounts whose tweets went viral in the context of this study. The anonymized version of the data used in this study is available upon request. %

\section{A Content Perspective}
\label{sec:contentperspective}

This section investigates the content characteristics of the tweets we gathered that discuss the war. 

\subsection{Hashtags}
\label{sec:hashtags}

\begin{figure*}
     \centering
     \subfloat[][]{
     \includegraphics[width=.5\linewidth, height=5cm]{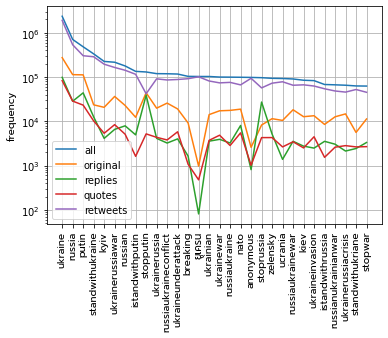} \label{lab:hashtags}}
     \subfloat[][]{\includegraphics[width=.5\linewidth, height=5cm]{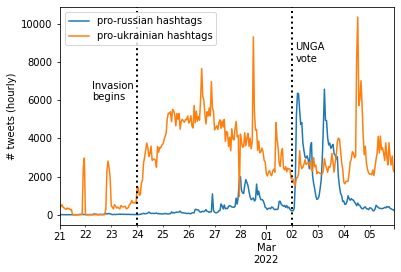}\label{lab:prohashtagstemporal}}
     \caption{(a) Top 30 hashtags along with their frequencies. (b) Number of tweets (per hour) containing pro-Russian vs. pro-Ukrainian hashtags.}
     \label{steady_state_2}
\end{figure*}

We start by inspecting the common hashtags encountered. We argue that these can provide insight into the nature of ongoing discussions, as well as the viewpoints of the participants.

\pb{Popular Hashtags.}
Figure~\ref{lab:hashtags} presents the frequency distribution of the top 30 hashtags. These hashtags are mostly stakeholder names that are directly involved or somehow affiliated with the war (\eg \#ukraine, \#russia, \#putin). 
Some other widely used hashtags express explicit solidarity with Ukraine (\eg \#standwithukraine, \#stopputin) while other show support for Russia (\eg \#istandwithputin, \#istandwithrussia). 
Apart from these hashtags, other highly used hashtags include \#anonymous, which refers to the announcement by the decentralized online hacking group Anonymous that they would be targeting Russian government websites and infrastructure. Interestingly, among the top hashtags, we also find a Thai language hashtag that translate to the word Ukraine.

Figure~\ref{lab:hashtags}
also shows the distribution of hashtags with respect to tweet types. Retweets amplify almost all hashtags except \#stopputin which is most frequently used in original tweets. 
This suggests that such tweets are largely generated by invested actors, but fail to gain wider traction via retweets.
Another contrasting trend is observed in the case of \#stoprussia, where replies surpass original tweets.
Such hashtags appear to engender lively discussions, leading to long threads.

To explore how these hashtags are used, we manually examine the 500 most frequent hashtags and identify the ones that show clear association with a certain stance (pro-Russia \vs pro-Ukraine). In total, we find 38 pro-Ukrainian hashtags that occur in 887,388 tweets, and 11 pro-Russian hashtags that are used in 218,417 tweets. We report these hashtags and their frequency of occurrence in Appendix A.

\pb{Temporal Distribution.}
Figure~\ref{lab:prohashtagstemporal} shows the temporal distribution of tweets containing pro-Ukrainian vs. pro-Russian hashtags. We observe contrasting differences in the amplification of hashtags supporting either of the two sides.
Specifically, the online narrative is tightly coupled with real-world events. The support for Ukraine became visible on the platform when \#StandWithUkraine started trending on 21st of February, \ie after the Russian president recognised two breakaway regions in Eastern Ukraine as independent entities. 
This support was further amplified on and after 24th of February (\ie the start of the invasion) with hashtags \#StandWithUkriane, \#IStandWithUkraine, \#StopRussia and \#StopPutin. 
In contrast, the first pair of pro-Russian hashtags (\#Hypocrisy \& \#AbolishNato) gained momentum between 27th and 28th of February after NATO condemned Russia’s full-scale invasion of Ukraine and said that it would deploy more troops to eastern Europe. Similarly, pro-Russian users used another pair of hashtags \ie \#IStandWithPutin \& \#IStandWithRussia to show solidarity with Russia when the United Nations General Assembly (UNGA) convened an emergency session to vote on resolution against Ukraine invasion.
Such hashtags shed clear light on international sentiment.

\subsection{Language \& Location}
\label{sec:language}
We conjecture that the above support might vary based on language and location.

\pb{Language.}
Twitter automatically assigns a BCP 47 language tag~\cite{wikipediaietflanguagetag} to each tweet. In line with our expectation, tweets in our dataset are predominantly English, as the majority of our search keywords are in English. Table~\ref{tab:languagestats} lists the top 10 languages. Clearly, Western languages dominate. We also report the percentage distribution of top 10 languages in pro-Ukrainian and pro-Russian tweets. English, being the most prevalent language overall, also dominate these subsets.

\begin{table*}[t]
\centering
\begin{tabular}{|l|l|lllll|ll|}
\hline
\multirow{2}{*}{\textbf{Language}} & \multicolumn{1}{c|}{\multirow{2}{*}{\textbf{Code}}} & \multicolumn{5}{c|}{\textbf{\# tweets}} & \multicolumn{2}{c|}{\textbf{Distribution (\%)}} \\ \cline{3-9} 
 & \multicolumn{1}{c|}{} & \multicolumn{1}{c|}{\textbf{original}} & \multicolumn{1}{c|}{\textbf{replies}} & \multicolumn{1}{c|}{\textbf{quotes}} & \multicolumn{1}{c|}{\textbf{retweets}} & \multicolumn{1}{c|}{\textit{\textbf{total}}} & \multicolumn{1}{c|}{\textbf{pro-Ukrainian}} & \multicolumn{1}{c|}{\textbf{pro-Russian}} \\ \hline
English & en & \multicolumn{1}{l|}{1,500,392} & \multicolumn{1}{l|}{2,344,185} & \multicolumn{1}{l|}{427,199} & \multicolumn{1}{l|}{15,559,215} & 19,830,991 & \multicolumn{1}{l|}{83.04} & 95.36 \\ \hline
Spanish & es & \multicolumn{1}{l|}{94,091} & \multicolumn{1}{l|}{122,184} & \multicolumn{1}{l|}{31,621} & \multicolumn{1}{l|}{909,770} & 1,157,666 & \multicolumn{1}{l|}{1.45} & 0.14 \\ \hline
French & fr & \multicolumn{1}{l|}{77,296} & \multicolumn{1}{l|}{75,318} & \multicolumn{1}{l|}{17,943} & \multicolumn{1}{l|}{631,080} & 801,637 & \multicolumn{1}{l|}{1.26} & 0.48 \\ \hline
Portuguese & pt & \multicolumn{1}{l|}{74,706} & \multicolumn{1}{l|}{98,447} & \multicolumn{1}{l|}{19,241} & \multicolumn{1}{l|}{352,295} & 544,689 & \multicolumn{1}{l|}{0.22} & 0.04 \\ \hline
German & de & \multicolumn{1}{l|}{68,176} & \multicolumn{1}{l|}{91,112} & \multicolumn{1}{l|}{13,114} & \multicolumn{1}{l|}{286,907} & 459,309 & \multicolumn{1}{l|}{2.24} & 0.53 \\ \hline
Undefined & und & \multicolumn{1}{l|}{93,643} & \multicolumn{1}{l|}{77,720} & \multicolumn{1}{l|}{38,095} & \multicolumn{1}{l|}{191,195} & 400,653 & \multicolumn{1}{l|}{4.23} & 1.40 \\ \hline
Italian & it & \multicolumn{1}{l|}{42,369} & \multicolumn{1}{l|}{44,953} & \multicolumn{1}{l|}{7,436} & \multicolumn{1}{l|}{180,595} & 275,353 & \multicolumn{1}{l|}{0.37} & 0.23 \\ \hline
Turkish & tr & \multicolumn{1}{l|}{39,093} & \multicolumn{1}{l|}{27,623} & \multicolumn{1}{l|}{7,201} & \multicolumn{1}{l|}{162,239} & 236,156 & \multicolumn{1}{l|}{0.99} & 0.03 \\ \hline
Japanese & ja & \multicolumn{1}{l|}{20,830} & \multicolumn{1}{l|}{4,864} & \multicolumn{1}{l|}{2,875} & \multicolumn{1}{l|}{176,905} & 205,474 & \multicolumn{1}{l|}{0.74} & 0.01 \\ \hline
Russian & ru & \multicolumn{1}{l|}{24,325} & \multicolumn{1}{l|}{37,727} & \multicolumn{1}{l|}{2,770} & \multicolumn{1}{l|}{135,386} & 200,208 & \multicolumn{1}{l|}{0.53} & 0.17 \\ \hline
\end{tabular}
\caption{Top 10 languages with their codes, number of tweets in the dataset, and percentage distribution in pro-Ukrainian \& pro-Russian tweets.}
\label{tab:languagestats}
\end{table*}

\begin{figure*}
     \centering
     \subfloat[][]{
     \includegraphics[width=.33\linewidth]{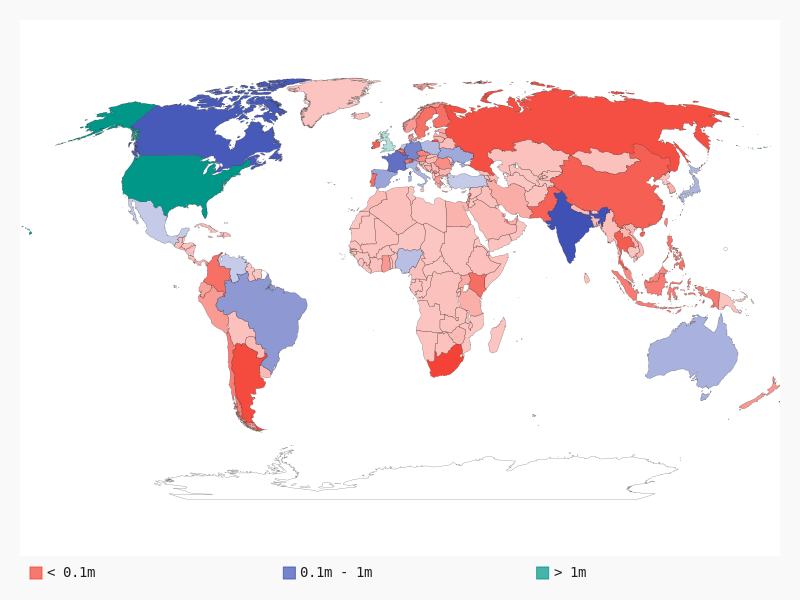} \label{lab:tweetslocation}}
     \subfloat[][]{\includegraphics[width=.33\linewidth]{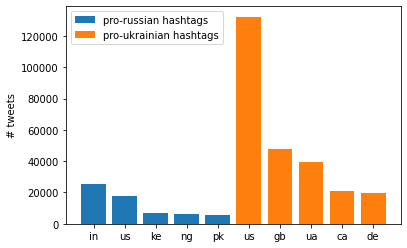}\label{lab:prohashtagslocation}}
     \subfloat[][]{\includegraphics[width=.33\linewidth]{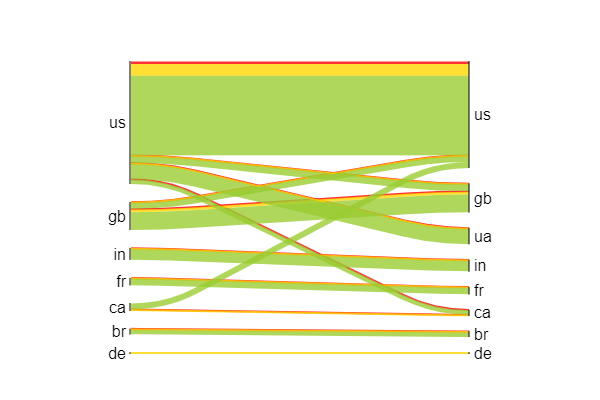}\label{lab:locationstats}}
     \caption{(a) Location of origin of tweets. (b) Location of origin of tweets containing pro-Ukrainian and pro-Russian hashtags. (c) Interaction flow (left to right) for retweets (yellow green), quotes (red) and replies (gold). The user from location on the left retweet, quote or reply a user from location on the right.}
     \label{steady_state_3}
\end{figure*}

\begin{figure*}[t]
     \centering
     \subfloat[][]{
     \includegraphics[width=.5\linewidth]{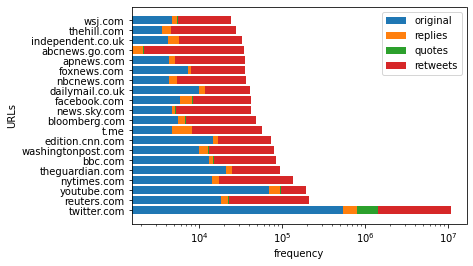} \label{lab:urls}}
     \subfloat[][]{\includegraphics[width=.5\linewidth]{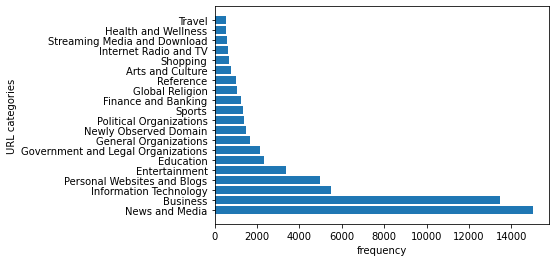}\label{lab:urlcategories}}
     \caption{(a) Top 20 URLs in the dataset. (b) Top 20 URL categories in the dataset.}
     \label{steady_state_4}
\end{figure*}

\pb{Locations.}
Figure~\ref{lab:tweetslocation} shows the location origin of tweets (based on their user location). The majority of tweets originate from Western countries with an exception of India, which ranks 3rd. 
Only a limited number of tweets in our dataset originate from Russia. This is naturally impacted by the censorship of Twitter in Russia. 
To better understand the types of regional support that exist, we inspect the location of tweets containing pro-Ukrainian vs. pro-Russian hashtags. Figure~\ref{lab:prohashtagslocation} shows the top 5 countries along with the number of tweets that originate from them for both subsets. This shows that the location of tweets supporting Ukraine is significantly different to that support Russia. Our results reveal that support for Ukraine mostly comes from the users in the global West whereas users from sub-continent and African countries are more inclined towards Russian narrative. We also briefly examine locations in scenarios where a user re-shares (retweets, quotes) or replies to other tweets. Figure~\ref{lab:locationstats} shows the interaction flow for retweets, quotes and replies based on location. The majority of time users interact with others in the same country. Interestingly, users from United States interact much more often with tweets originating from Ukraine than that from Russia. 

\subsection{URL Sharing}
\label{sec:urls}

We next turn to the shared URLs to explore what type of content that is posted related to the war. We observe nearly 2.4 million unique URLs (80,815 unique base URLs), which are shared a total of 15 million times. 
Figure~\ref{lab:urls} lists the top 20 domains posted. Most of the widely used domains are either mainstream news websites (\eg Reuters, New York Times) or social media platforms (\eg YouTube, Facebook). 
We also find t.me URLs among the top 10 most posted domains. These URLs correspond to the popular messaging platform, Telegram, which became the go-to messaging app for Ukrainian due to its unique features such as `secret chat'~\cite{telegraminukraine}.

To better understand the nature of these tweets, we classify the domains shared using FortiGuard~\cite{fortiguardapi}. We identify 86 different domain categories with 23\% domains belonging to `News and Media`. Figure~\ref{lab:urlcategories} presents the top 20 URL categories in the dataset. Interestingly, we also find a small number of `Pornography` (327), `Phishing` (202) and `Spam` (53) URLs.
Such users appear to be exploiting the viral popularity of Ukrainian posts to promote these links.

\subsection{Toxicity Analysis}
\label{sec:toxicity}

\begin{figure*}
     \centering
     \subfloat[][]{
     \includegraphics[width=.5\linewidth, height=5cm]{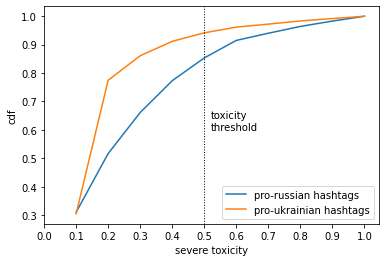} \label{lab:prorussianvsproukrainiantoxicity}}
     \subfloat[][]{\includegraphics[width=.5\linewidth, height=5cm]{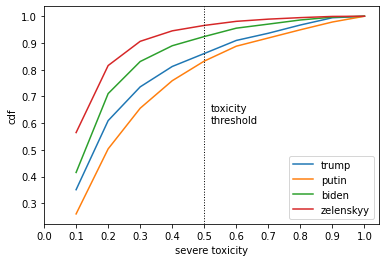}\label{lab:toxicitypresidents}}
     \caption{(a) CDF of the toxicity scores for the tweets containing pro-Russian and pro-Ukrainian hashtags. (b) CDF of the toxicity scores for the tweets mentioning Zelenskyy, Biden, Trump or Putin.}
     \label{steady_state_5}
\end{figure*}

Polarization and toxicity in political discourse is common on Twitter~\cite{saveski2020polarization}.
Figure~\ref{lab:prorussianvsproukrainiantoxicity} shows the Cumulative Distribution Function (CDF) of the Perspective toxicity scores for the tweets containing pro-Russian and pro-Ukrainian hashtags. In the literature, 0.5 is the most common choice to threshold the perspective scores~\cite{bin2022toxicity,rottger2020hatecheck,papasavva2020raiders}, however, higher values such as 0.8 are also used~\cite{agarwal2021hate}. Here, we use 0.5 as a threshold and consider a tweet to be toxic if its toxicity score is greater than 0.5 (and vice versa). 

Given this threshold, we find that 14.7\% of the tweets containing pro-Russian hashtags are toxic compared to just 5.9\% of pro-Ukrainian tweets. The toxic tweets with pro-Russian hashtags are mostly anti-US/NATO and cite the hypocrisy of the West over conflicts in Palestine, Syria and Iraq (using swear words). 
In contrast, the toxicity in the pro-Ukrainian tweets is mostly directed towards Russian president Putin and former US president Trump. In fact, among the tweets mentioning important political figures the tweets mentioning Trump has the second highest toxicity percentage (Figure~\ref{lab:toxicitypresidents}), even though the invasion took place after Trump left office. 
Closer inspection suggests that Trump attracted this backlash after calling Putin a genius.\footnote{https://edition.cnn.com/2022/03/03/politics/trump-putin-russia-ukraine-graham/index.html}
We argue that such insights can be vital for understanding the wider public reaction to political figures and policies.

\subsection{Withheld Tweets}
\label{sec:witheld}

\begin{figure}[t]
     \centering
     \includegraphics[width=0.9\linewidth]{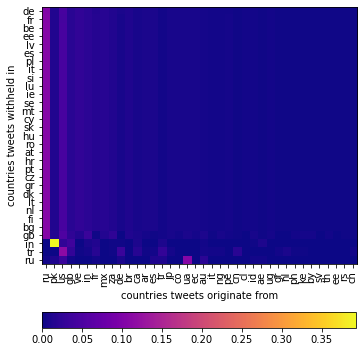}
     \caption{Fraction of tweets withheld in a country on y-axis that originate from a country on x-axis.}
     \label{lab:withheldheatmap}
\end{figure}

\begin{figure*}
     \centering
     \subfloat[][]{
     \includegraphics[width=.5\linewidth, height=5cm]{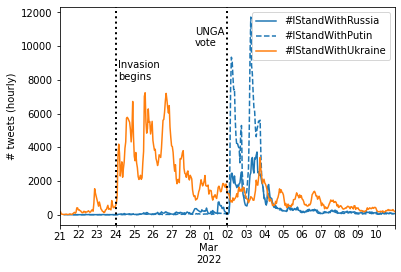} \label{lab:tweetsprorussiaproukraine}}
     \subfloat[][]{\includegraphics[width=.5\linewidth, height=5cm]{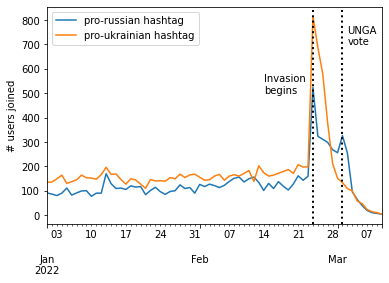}\label{lab:newusersprorussiaproukraine}}
     \caption{(a) Number of tweets (per hour) containing \#IStandWithPutin, \#IStandWithRussia and \#IStandWithUkraine hashtags. (b) Number of users joined (daily) since January 01, 2022 till March 10, 2022 in pro-Russian and pro-Ukrainian subsets.}
     \label{steady_state_6}
\end{figure*}

The above level of toxicity leads us to conjecture that relevant tweets may be withheld in certain jurisdictions. Note, this is where the state authority requires Twitter to prevent such tweets from being seen by users in that country.\footnote{See https://help.twitter.com/en/rules-and-policies/tweet-withheld-by-country}

In total, 56,512 tweets (0.22\% of the tweets in our dataset) are withheld across 33 different countries. These tweets are collectively posted by 25,733 users, out of which 113 are verified and 448 have their profile withheld in at least one country. 
Figure~\ref{lab:withheldheatmap} shows the fraction of tweets withheld in a country (y-axis) that originate from each country (x-axis) as a heatmap. Here, we only plot countries that withhold at least 100 tweets or at least 100 tweets originating from them has been withheld. Countries in the European Union and United Kingdom withhold the most tweets and the largest fraction of these tweets originate from Russia. Some of the prominent Russian accounts whose tweets are withheld are state-owned media \eg @RTarabic and @RT\_russian. 
These accounts primarily share news articles about the Russia-Ukraine war in their tweets.\footnote{Note, evaluating the authenticity of these news articles is beyond the scope of this work}
Unsurprisingly, 10.6\% of tweets withheld in Russia originate from Ukraine (note, in March, Russia blocked all access to Twitter~\cite{guardianrussiablockstwitter}). 
The tweets withheld in Russia are from accounts that are very vocal about Russian Invasion of Ukraine, \eg @ChechenCenter - an independent media from Chechen Republic of Ichkeria. Since Twitter does not provide the date a tweet is withheld, it is hard to say how many tweets are withheld as a consequence of Russia-Ukraine conflict. However, it is evident that not all of the tweets are withheld in the aftermath of Russia-Ukraine war, \eg we observe a large fraction of tweets withheld in India that originate from Pakistan --- this is simply because of the usual political tensions between the two countries~\cite{elmas2021dataset}. 

\section{A User Perspective}
\label{sec:userperspetive}

The previous section has shown that there is a wide range of both pro-Russian and pro-Ukranian discourse, with a unusually high prevalence of toxic text. We next proceed to explore the users sharing this content.
To understand, the partisan nature of the users, we extract all users who have used the 
pro-Russian \#IStandWithPutin and \#IStandWithRussia hashtags;
then all the users who have posted on the \#IStandWithUkraine hashtag.
To expand this dataset, we further gather all the tweets posted between February 21, 2022 and March 10, 2022 containing these three hashtags. Figure~\ref{lab:tweetsprorussiaproukraine} plots the count of tweets for each hashtag. 
In total, we collect 260,138 tweets with \#IStandWithPutin, 149,106 with \#IStandWithRussia and 638,217 with \#IStandWithUkraine.

\subsection{New Users}
\label{sec:newusers}

\begin{figure*}
     \centering
     \subfloat[][]{
     \includegraphics[width=.33\linewidth]{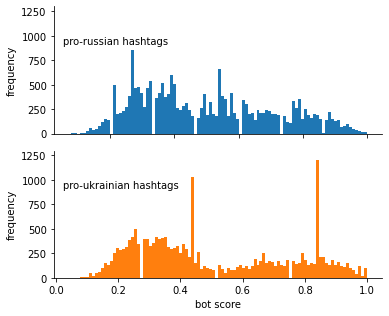} \label{lab:botometerdistribution}}
     \subfloat[][]{\includegraphics[width=.33\linewidth]{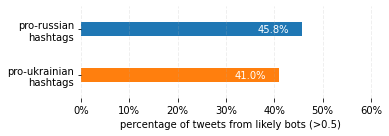}\label{lab:botometerdistributionpercentage0.5}}
     \subfloat[][]{\includegraphics[width=.33\linewidth]{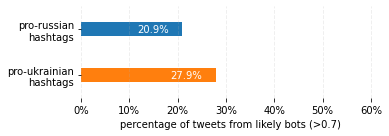}\label{lab:botometerdistributionpercentage0.7}}
     \caption{(a) Bot score distribution for tweets from both pro-Russian and pro-Ukrainian subsets. (b) Percentage of tweets posted by likely bots using 0.5 as a threshold. (c) Percentage of tweets posted by likely bots using 0.7 as a threshold.}
     \label{lab:botometergraphs}
\end{figure*}

We first inspect the number of new users who contribute to online discussions. 
Tweets in the pro-Russian subset (\ie containing \#IStandWithPutin \& \#IStandWithRussia) are posted by 120,370 unique users, whereas the ones in pro-Ukrainian subset (\ie containing \#IStandWithUkraine) are posted by 298,867 unique users. We conjecture that new users are more likely to be inauthentic as they may be put in place just to coordinate campaigns and amplify narratives.

To explore this, we look at the creation dates of users in the pro-subsets. Figure~\ref{lab:newusersprorussiaproukraine} shows the number of users who joined (daily) since January 01, 2022 till March 10, 2022 in pro-Russian and pro-Ukrainian subsets. The pro-Russian subset contains 9,008 users who joined between January 01, 2022 and March 10, 2022. These users collectively posted 30,343 tweets. In comparison, pro-Ukrainian subset contains 11,854 such users who posted a total of 25,284 tweets. 

Interestingly, we observe a large spike in the number of new users on and after February 24, 2022 (\ie in the early days of invasion) followed by a decreasing trend after a week, in both subsets. Nearly 132\% more users joined in the week after the invasion compared to the week before it, collectively in both subsets. 
Curiously, opposed to one large peak in the pro-Ukrainian subset, we observe two dramatic peaks in the pro-Russian users: one at the beginning of the invasion, and the other on the day of UN General Assembly session. 
This influx could be driven by genuine social media users interested in the public discourse, just as the subsequent decreasing trend could be due to a decline in public interest or Russia's ban on Twitter (March 04, 2022~\cite{guardianrussiablockstwitter}).
However, we conjecture that these patterns could also be driven by an influx of coordinated accounts (\eg bots) driven by state actors.

\subsection{Presence of Bots}
\label{sec:bots}

To see whether these new users are bot-like, we pass all the users created on or after January 01, 2022 through Botometer~\cite{davis2016botornot}.\footnote{Botometer is a supervised machine learning classifier that distinguishes bot-like from human-like accounts based on their 1000+ features \eg length of screen name, profile picture and background, age of account etc.} 
For each user account, Botometer provides a ``bot score`` between 0 and 1: a score close to 1 means the account is highly automated, while a score near 0 means a human is likely handling the account. 

We plot the bot score distribution for tweets posted by new users from both pro-Russian and pro-Ukrainian subsets in Figure~\ref{lab:botometerdistribution}. Since we are interested in the bot activity level of pro-hashtags, we use tweets (as opposed to accounts) as the units of analysis. 
In both the subsets, the distribution has a bimodal pattern. 
The spikes appear to be driven by accounts tweeting the same hashtag repeatedly. For example, the spike at 0.84 in the pro-Ukrainian subset comes from a bot account that tweets ``F**k you, Putin!`` every minute in various languages.

While this particular parody bot account is honest about itself, the majority of the bots are not. Thus, to calculate the proportion of tweets from likely bots in both pro-subsets, we consider accounts with
scores higher than a threshold as likely bots. In the literature, 0.5 is the most common choice of threshold~\cite{shao2018spread,bessi2016social,vosoughi2018spread}, however, higher values, such as 0.7~\cite{grinberg2019fake} and 0.8~\cite{broniatowski2018weaponized}, are also used. We use both 0.5 and 0.7 as thresholds and show the results in Figure~\ref{lab:botometerdistributionpercentage0.5}
and \ref{lab:botometerdistributionpercentage0.7} respectively.

These results confirm a significant presence of bots on \emph{both} sides, albeit with pro-Russian hashtag posts more in number.
We apply two-proportions $z$-tests to estimate the significance level of the differences. When using 0.5 as the threshold, the percentage of tweets from likely bots that mention pro-Russian hashtags is significantly higher than those that use pro-Ukrainian hashtags ($p < 0.01$). Interestingly, the results reverse when we apply 0.7 threshold \ie the percentage of tweets from likely bots in the pro-Russian subset becomes significantly lower than that in pro-Ukrainian subset ($p < 0.01$). If we revisit Figure~\ref{lab:botometerdistribution}, we see that although the bot score distribution of pro-Russian hashtags has more mass in the (0.5, 1] region than the distribution of pro-Ukrainian hashtags, the mass tends to concentrate around 0.6.
In contrast, the distribution of pro-Ukrainian hashtags has more mass around 0.85. This nuanced difference causes the contradictory results when using different thresholds.
Put simply, new users that tweeted in support of Russia show more bot-like activity.
Yet the pro-Ukranian bots exhibit more obvious traits. 
This stands in contrast to anecdotal popular opinion and we conjecture that this may be driven by greater sophistication among the pro-Russian bots.

\subsection{User Interactions}
\label{sec:interactiongraph}
We conjecture that the above bots may drive certain patterns in the retweet network.
Thus, to investigate any attempts at platform manipulation~\cite{twittermanipulationpolicy} in building or amplifying a certain narrative, we next look into the most influential tweets in both pro-subsets.

\begin{figure*}
     \centering
     \subfloat[][]{
     \includegraphics[width=.5\linewidth, height=5cm]{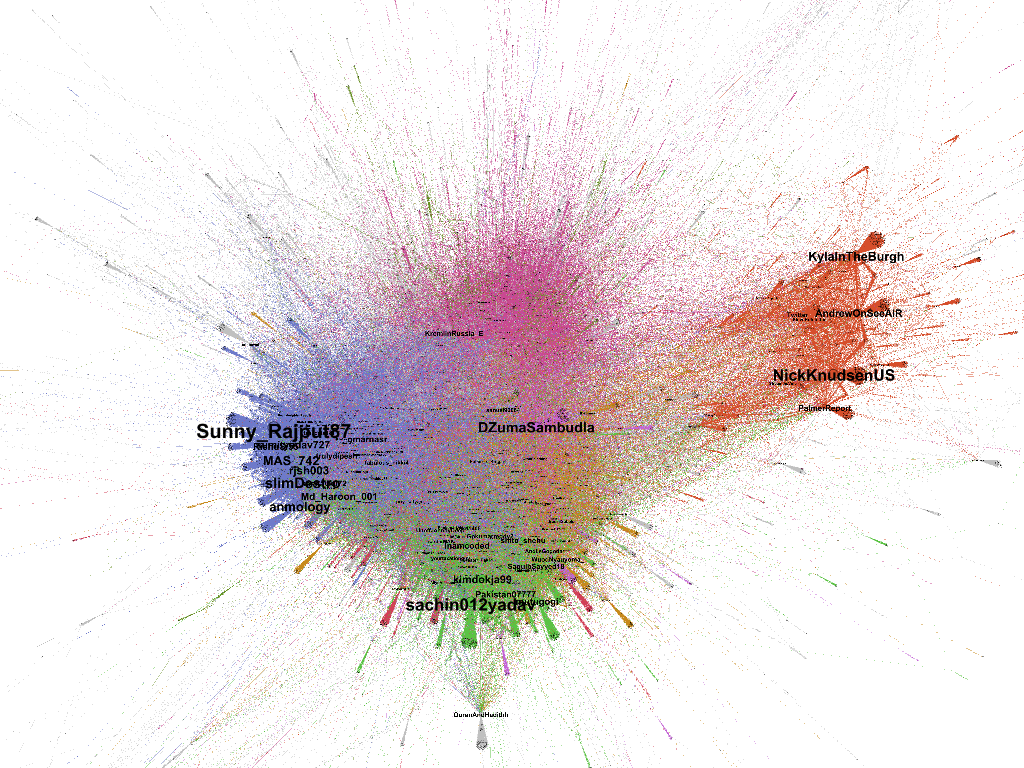} \label{lab:prorussia}}
     \subfloat[][]{\includegraphics[width=.5\linewidth, height=5cm]{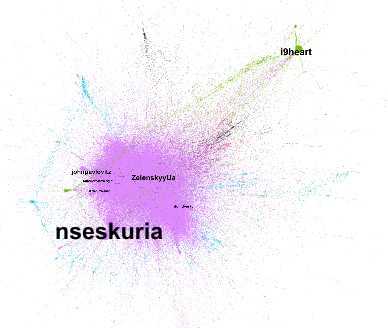}\label{lab:proukraine}}
     \caption{Interaction graph of users in both (a) pro-Russian and (b) pro-Ukrainian subsets. Each node represents a user and each edge represents a retweet directed from a retweeting to retweeted user.}
     \label{lab:retweetnetworks}
\end{figure*}

\pb{Community Exploration.}
Figure~\ref{lab:retweetnetworks} presents the interaction graph of users in both pro-Russian (Figure~\ref{lab:prorussia}) and pro-Ukrainian (Figure~\ref{lab:proukraine}) subsets. Here, each node represents a user and each edge represents a retweet directed from a retweeting to retweeted user. The node and label sizes are directly proportional to the number of retweets \ie the higher the number of retweets a user receives the bigger is its node and label. Each colour represent a different community \ie the users that tend to interact with each other.

\pb{Pro-Russian Patterns.}
In case of the pro-Russian subset, we observe two distinct groups. The first group (left in Figure~\ref{lab:prorussia}) consists of many dense communities that use the pro-Russian hashtags to support the Russian narrative whereas the second cluster (right) consists of a single community that used the pro-Russian hashtags, but to condemn the Russian cause. 
This highlights the challenges of using the hashtags to assign stance to the users --- many people include supportive hashtags from the opposing side as part of their critique.
The most retweeted account from the first cluster is @Sunny\_Rajput --- a relatively new account (joining in September 2021) of Indian origin but claims itself to be from United States (396 followers and follows 1,301).
The tweet posted by this account received 5,171 retweets. Other popular tweets in this cluster are from @slimDestro (joined: March 2011, from: India, followers: 102, follows: 5, retweets received: 3,811), @sachin012yadav (joined: December 2015, from: India, followers: 74, follows: 90, retweets received: 3,630) and @anmology (joined: October 2018, from: India, followers: 145, follows: 50, retweets received: 3,362). Interestingly, top 10 most retweeted tweets in this cluster received suspiciously large number of retweets than the number of followers of accounts who posted them, raising questions over the virality of these tweets. The tweets posted by these accounts centre around Western hypocrisy, double standards and support for Putin. 

The right cluster in Figure~\ref{lab:prorussia} is rather different, with the most viral tweet coming from @NickKnudsenUS.
This user joined in November 2016 from United States, has 229,517 followers, follows 19,413 and received 4,064 retweets.
In contrast to the pro-Russian tweets in the left cluster, this one is actually using the pro-Russian hashtag \#IStandWithPutin to condemn the war, calling Putin's supporters fascists. This again highlights the challenge of deriving stance from the use of hashtags alone.

\pb{Pro-Ukranian Patterns.}
The interaction graph of users using pro-Ukrainian hashtags is rather different, with fewer clear communities of users emerging.
This suggests that pro-Ukranian prominent voices are consolidated around a much small number of key players.
Users here show solidarity with Ukraine, raise voices for aid, express anger towards Trump, and call Putin a war criminal. 
The most retweeted tweet here is from @nseskuria, who posted a photo of the Mayor of Kyiv in the war zone. 
This account joined in October 2014, originating from the United Kingdom, has 8,211 followers, follows 2,002 and received 23,976 retweets. 
Manual inspection confirms this is not a bot. 
Other popular users in the pro-Ukrainian subset include @i9heart (joined: December 2021, from: N/A, followers: 232, follows: 210, retweets received: 15,458), @johnpavlovitz (joined: February 2012, from: United States, followers: 382,191, follows: 327,69, retweets received: 11,224) and @MuellerSheWrote (joined: November 2017, from: United States, followers: 241,619, follows: 8,378, retweets received: 6,634). 
We argue that these patterns can support further analysis to understand the coordinated spread of posts.

\section{Related Work} \label{sec:related_work}

\pb{Studies of Real-World Events.}
There have been a number of prior studies looking at real-world events through the lens of social media~\cite{lumezanu2012bias}.
Social media has been studied as an effective tool for factual reporting and situational awareness~\cite{ahmed2019digital}. Social media has also been used to study opinion engineering~\cite{solo2017overview}, political polarization~\cite{haq2020enemy,lumezanu2012bias}, and propaganda campaigns~\cite{keller2017manipulate} in social events.

\pb{Social Media \& War.}
The Russia-Ukraine war has already been the focus of several studies~\cite{gutmann2022has,ibar2022opinion,caprolu2022characterizing}.
These have had a particular emphasis on  information quality~\cite{pierri2022propaganda}.
Similar to our own work, both Ukrainian and Russian supporters have been part of information warfare across various social media and online platforms~\cite{chen2022tweets,haq2022weapon}. 
One of our key findings is that countries are withholding tweets.
A recent comparison of Twitter and Facebook found that most accounts involved in propaganda are politically right-leaning. However, content moderation on such platforms could only remove up to 15\% of such content~\cite{pierri2022propaganda}. 
Our results also show that many new accounts participate in discussions related to the war. Pierri et al. studied banned accounts during the conflict.
They found that such accounts' creation and suspension dates are clustered around temporal epochs~\cite{pierri2022does}.
They also found that such accounts primarily interact with normal users through replies and mentions. 

Besides censorship, selective media reporting has also been observed widely. Conflict coverage from Russian state-affiliated media has been found to differ from independent media; reducing conflict situational awareness to the Russian population and leaving them more vulnerable to disinformation~\cite{park2022voynaslov}. The race to share news quickly also compromises the quality.
For example, fake videos from platforms like TikTok have been shared without fact-checking from several media outlets~\cite{stuanescu2022ukraine}. 
Another study has shown that users who report fake news exhibit ``motivated reasoning''. This study found that Ukrainian users are more like to endorse fake news regarding Russia as compared to fake news for the European Union~\cite{osmundsen2022information}.

Other systematic ways to spread misinformation and propaganda are also identified during the war. One study found that most of pro-Russia campaigns were geographically distributed and about 20\% of content came from the bots created during the earlier days of the crisis~\cite{geissler2022russian}. In addition, the use of social issues and polarization, such as \textit{Gun Control} in the US, is also highlighted as one of the prominent way to spread the online disinformation during the war~\cite{yablokov2022russian}.

Others have measured public sentiment related to the war via social media data~\cite{ngo2022public}. 
In-line with our findings, they show that support varies based on region, with users from Western countries being most supportive of the sanctions on Russia. 
Interestingly, Twitter sentiment also correlated with Forex rates and crude oil prices~\cite{polyzos2022escalating}.

\pb{Novelty.}
Our work supplements previous works on Twitter discourse during the conflict.
We particularly highlight the posting across partisan hashtags, differences among regions, and censorship by different countries. In addition, we highlight the role of automated accounts to amplify certain narratives.

\section{Conclusion}

This paper has presented a study of Twitter discourse surrounding the Russian invasion of Ukraine. 
We wish to highlight a set of key findings.
We found that pro-Ukrainian content is more prevalent on Twitter than pro-Russian content, and that the online discourse is tightly coupled with the real-world political events (\S\ref{sec:hashtags}).
We observed differences across regions and languages (\S\ref{sec:language}), with, Western narratives dominating the dataset. Whereas Ukraine gathers support from the global West, users in the sub-continent and Africa are more inclined towards the Russian narrative.
Toxicity among these discussions is commonplace, particularly for pro-Russian hashtags (\S\ref{sec:toxicity}). 
For example, 14.7\% of the tweets containing pro-Russian hashtags are toxic compared to just 5.9\% of pro-Ukrainian tweets.
This leads to tweets being withheld in certain countries (\S\ref{sec:witheld}).
We found that countries in Europe withhold the most tweets, and the largest fraction of these tweets originate from Russia. A significant surge of posts come from new users (\S\ref{sec:newusers}) and in-line with anecdotal evidence, we find that a large number of pro-Russian tweets are driven by bots (\S\ref{sec:bots}). 

We emphasise that this work constitutes just the first step in exploring the online discourse surrounding the war.
We wish to highlight two key limitations that we aim to address in our future work. 
First, our dataset is naturally biased towards Western perspectives, particularly as Russia blocked Twitter in March 2022. To address this, we plan to perform similar analysis on more pro-Russian social networks such as VK.
Second, we found significant noise when using hashtags to assign stance to individual tweets. For instance, many pro-Russian hashtags were included in pro-Ukrainian tweets (as part of a critique). Consequently, the above analysis should be interpreted with this in-mind. 
To address this, we hope to build more robust classifiers that can label tweets as pro- Ukraine or Russian, without exclusive reliance on hashtags.
We argue that this is vital as social media will increasingly become an arena for real-world conflicts.

\bibliographystyle{plain}
\bibliography{sample-base, ref}

\begin{thebibliography}{10}

\bibitem{agarwal2021hate}
Pushkal Agarwal, Oliver Hawkins, Margarita Amaxopoulou, Noel Dempsey, Nishanth
  Sastry, and Edward Wood.
\newblock Hate speech in political discourse: A case study of uk mps on
  twitter.
\newblock In {\em Proceedings of the 32nd ACM Conference on Hypertext and
  Social Media}, pages 5--16, 2021.

\bibitem{ahmed2019digital}
Nova Ahmed, Zareen Tasnim, and Jasmine Jones.
\newblock Digital silence and liberating stories: during a student-driven
  movement.
\newblock In {\em Extended Abstracts of the 2019 CHI Conference on Human
  Factors in Computing Systems}, pages 1--10, 2019.

\bibitem{telegraminukraine}
Mamoun Alazab and Kate Macfarlane.
\newblock Why telegram became the go-to app for ukrainians – despite being
  rife with russian disinformation.
\newblock
  \url{https://theconversation.com/why-telegram-became-the-go-to-app-for-ukrainians-despite-being-rife-with-russian-disinformation-179560},
  2022.
\newblock Accessed: 2023-01-13.

\bibitem{bessi2016social}
Alessandro Bessi and Emilio Ferrara.
\newblock Social bots distort the 2016 us presidential election online
  discussion.
\newblock {\em First monday}, 21(11-7), 2016.

\bibitem{bin2022toxicity}
Haris Bin~Zia, Aravindh Raman, Ignacio Castro, Ishaku Hassan~Anaobi, Emiliano
  De~Cristofaro, Nishanth Sastry, and Gareth Tyson.
\newblock Toxicity in the decentralized web and the potential for model
  sharing.
\newblock {\em Proceedings of the ACM on Measurement and Analysis of Computing
  Systems}, 6(2):1--25, 2022.

\bibitem{broniatowski2018weaponized}
David~A Broniatowski, Amelia~M Jamison, SiHua Qi, Lulwah AlKulaib, Tao Chen,
  Adrian Benton, Sandra~C Quinn, and Mark Dredze.
\newblock Weaponized health communication: Twitter bots and russian trolls
  amplify the vaccine debate.
\newblock {\em American journal of public health}, 108(10):1378--1384, 2018.

\bibitem{caprolu2022characterizing}
Maurantonio Caprolu, Alireza Sadighian, and Roberto Di~Pietro.
\newblock Characterizing the 2022 russo-ukrainian conflict through the lenses
  of aspect-based sentiment analysis: Dataset, methodology, and preliminary
  findings.
\newblock {\em arXiv preprint arXiv:2208.04903}, 2022.

\bibitem{chen2022tweets}
Emily Chen and Emilio Ferrara.
\newblock Tweets in time of conflict: A public dataset tracking the twitter
  discourse on the war between ukraine and russia.
\newblock {\em arXiv preprint arXiv:2203.07488}, 2022.

\bibitem{davis2016botornot}
Clayton~Allen Davis, Onur Varol, Emilio Ferrara, Alessandro Flammini, and
  Filippo Menczer.
\newblock Botornot: A system to evaluate social bots.
\newblock In {\em Proceedings of the 25th international conference companion on
  world wide web}, pages 273--274, 2016.

\bibitem{elmas2021dataset}
Tugrulcan Elmas, Rebekah Overdorf, and Karl Aberer.
\newblock A dataset of state-censored tweets.
\newblock In {\em ICWSM}, pages 1009--1015, 2021.

\bibitem{geissler2022russian}
Dominique Geissler, Dominik B{\"a}r, Nicolas Pr{\"o}llochs, and Stefan
  Feuerriegel.
\newblock Russian propaganda on social media during the 2022 invasion of
  ukraine.
\newblock {\em arXiv preprint arXiv:2211.04154}, 2022.

\bibitem{grinberg2019fake}
Nir Grinberg, Kenneth Joseph, Lisa Friedland, Briony Swire-Thompson, and David
  Lazer.
\newblock Fake news on twitter during the 2016 us presidential election.
\newblock {\em Science}, 363(6425):374--378, 2019.

\bibitem{guardianrussiablockstwitter}
Guardian.
\newblock Russia blocks access to facebook and twitter.
\newblock
  \url{https://www.theguardian.com/world/2022/mar/04/russia-completely-blocks-access-to-facebook-and-twitter},
  2022.
\newblock Accessed: 2022-07-27.

\bibitem{gutmann2022has}
Jerg Gutmann, Hans Pitlik, and Andrea Fronasch{\"u}tz.
\newblock Has the russian invasion of ukraine reinforced anti-globalization
  sentiment in austria?
\newblock 2022.

\bibitem{haq2020enemy}
Ehsan~ul Haq, Tristan Braud, Young~D Kwon, and Pan Hui.
\newblock Enemy at the gate: evolution of twitter user's polarization during
  national crisis.
\newblock In {\em Proceedings of the 12th IEEE/ACM International Conference on
  Advances in Social Networks Analysis and Mining}, pages 212--216, 2020.

\bibitem{haq2022weapon}
Ehsan-Ul Haq, Gareth Tyson, Tristan Braud, and Pan Hui.
\newblock Weaponising social media for information divide and warfare.
\newblock In {\em Proceedings of the 33rd ACM Conference on Hypertext and
  Social Media}, HT '22, page 259–262, New York, NY, USA, 2022. Association
  for Computing Machinery.

\bibitem{ibar2022opinion}
Raquel Ibar-Alonso, Raquel Quiroga-Garc{\'\i}a, and Mar Arenas-Parra.
\newblock Opinion mining of green energy sentiment: A russia-ukraine conflict
  analysis.
\newblock {\em Mathematics}, 10(14):2532, 2022.

\bibitem{jiang2020political}
Julie Jiang, Emily Chen, Shen Yan, Kristina Lerman, and Emilio Ferrara.
\newblock Political polarization drives online conversations about covid-19 in
  the united states.
\newblock {\em Human Behavior and Emerging Technologies}, 2(3):200--211, 2020.

\bibitem{keller2017manipulate}
Franziska~B Keller, David Schoch, Sebastian Stier, and JungHwan Yang.
\newblock How to manipulate social media: Analyzing political astroturfing
  using ground truth data from south korea.
\newblock In {\em Eleventh international AAAI conference on Web and Social
  Media}, 2017.

\bibitem{fortiguardapi}
FortiGuard Labs.
\newblock Web filter lookup.
\newblock \url{https://www.fortiguard.com/webfilter?q=<URL>&version=9}, 2022.
\newblock Accessed: 2022-07-27.

\bibitem{lumezanu2012bias}
Cristian Lumezanu, Nick Feamster, and Hans Klein.
\newblock \# bias: Measuring the tweeting behavior of propagandists.
\newblock In {\em Proceedings of the International AAAI Conference on Web and
  Social Media}, volume~6, pages 210--217, 2012.

\bibitem{ngo2022public}
Vu~M Ngo, Toan~LD Huynh, Phuc~V Nguyen, Huan~H Nguyen, et~al.
\newblock Public sentiment towards economic sanctions in the russia-ukraine
  war.
\newblock Technical report, Global Labor Organization (GLO), 2022.

\bibitem{openstreetmapsearchapi}
Nominatim.
\newblock Openstreetmap search api.
\newblock \url{https://nominatim.openstreetmap.org/search?<params>}, 2022.
\newblock Accessed: 2022-08-22.

\bibitem{osmundsen2022information}
Mathias Osmundsen, Michael~Bang Petersen, Honorata Mazepus, Dimiter Toshkov,
  and Antoaneta Dimitrova.
\newblock Information battleground: Conflict perceptions motivate the belief in
  and sharing of fake news about the adversary.
\newblock 2022.

\bibitem{papasavva2020raiders}
Antonis Papasavva, Savvas Zannettou, Emiliano De~Cristofaro, Gianluca
  Stringhini, and Jeremy Blackburn.
\newblock Raiders of the lost kek: 3.5 years of augmented 4chan posts from the
  politically incorrect board.
\newblock In {\em Proceedings of the International AAAI Conference on Web and
  Social Media}, volume~14, pages 885--894, 2020.

\bibitem{park2022voynaslov}
Chan~Young Park, Julia Mendelsohn, Anjalie Field, and Yulia Tsvetkov.
\newblock Voynaslov: A data set of russian social media activity during the
  2022 ukraine-russia war.
\newblock {\em arXiv preprint arXiv:2205.12382}, 2022.

\bibitem{perspectiveapi}
Perspective.
\newblock Perspective api.
\newblock \url{https://www.perspectiveapi.com}, 2022.
\newblock Accessed: 2022-08-22.

\bibitem{pierri2022does}
Francesco Pierri, Luca Luceri, and Emilio Ferrara.
\newblock How does twitter account moderation work? dynamics of account
  creation and suspension during major geopolitical events.
\newblock {\em arXiv preprint arXiv:2209.07614}, 2022.

\bibitem{pierri2022propaganda}
Francesco Pierri, Luca Luceri, Nikhil Jindal, and Emilio Ferrara.
\newblock Propaganda and misinformation on facebook and twitter during the
  russian invasion of ukraine.
\newblock {\em arXiv preprint arXiv:2212.00419}, 2022.

\bibitem{polyzos2022escalating}
Efstathios~Stathis Polyzos.
\newblock Escalating tension and the war in ukraine: Evidence using impulse
  response functions on economic indicators and twitter sentiment.
\newblock {\em Available at SSRN 4058364}, 2022.

\bibitem{rottger2020hatecheck}
Paul R{\"o}ttger, Bertram Vidgen, Dong Nguyen, Zeerak Waseem, Helen Margetts,
  and Janet Pierrehumbert.
\newblock Hatecheck: Functional tests for hate speech detection models.
\newblock {\em arXiv preprint arXiv:2012.15606}, 2020.

\bibitem{saveski2020polarization}
Martin Saveski.
\newblock {\em Polarization and toxicity in political discourse online}.
\newblock PhD thesis, Massachusetts Institute of Technology, 2020.

\bibitem{shao2018spread}
Chengcheng Shao, Giovanni~Luca Ciampaglia, Onur Varol, Kai-Cheng Yang,
  Alessandro Flammini, and Filippo Menczer.
\newblock The spread of low-credibility content by social bots.
\newblock {\em Nature communications}, 9(1):1--9, 2018.

\bibitem{solo2017overview}
Ashu~MG Solo.
\newblock An overview of the new interdisciplinary fields of political
  engineering and computational politics for the next frontier in politics.
\newblock In {\em 2017 International Conference on Computational Science and
  Computational Intelligence (CSCI)}, pages 1805--1806. IEEE, 2017.

\bibitem{stuanescu2022ukraine}
Georgiana St{\u{a}}nescu.
\newblock Ukraine conflict: the challenge of informational war.
\newblock 2022.

\bibitem{twittermanipulationpolicy}
Twitter.
\newblock Platform manipulation and spam policy.
\newblock
  \url{https://help.twitter.com/en/rules-and-policies/platform-manipulation},
  2022.
\newblock Accessed: 2022-07-27.

\bibitem{twittersearchapi}
Twitter.
\newblock Twitter search api.
\newblock
  \url{https://developer.twitter.com/en/docs/twitter-api/tweets/search/introduction},
  2022.
\newblock Accessed: 2022-08-22.

\bibitem{vosoughi2018spread}
Soroush Vosoughi, Deb Roy, and Sinan Aral.
\newblock The spread of true and false news online.
\newblock {\em science}, 359(6380):1146--1151, 2018.

\bibitem{wikipediacrimes}
Wikipedia.
\newblock Annexation of crimea by the russian federation.
\newblock
  \url{https://en.wikipedia.org/wiki/Annexation_of_Crimea_by_the_Russian_Federation},
  2014.
\newblock Accessed: 2022-08-22.

\bibitem{wikipediaietflanguagetag}
Wikipedia.
\newblock Ietf language tag.
\newblock \url{https://en.wikipedia.org/wiki/IETF_language_tag}, 2022.
\newblock Accessed: 2022-08-22.

\bibitem{wikipediukraine}
Wikipedia.
\newblock Russian invasion of ukraine.
\newblock \url{https://en.wikipedia.org/wiki/Russian_invasion_of_Ukraine},
  2022.
\newblock Accessed: 2022-08-22.

\bibitem{yablokov2022russian}
Ilya Yablokov.
\newblock Russian disinformation finds fertile ground in the west.
\newblock {\em Nature Human Behaviour}, pages 1--2, 2022.

\bibitem{zannettou2019disinformation}
Savvas Zannettou, Tristan Caulfield, Emiliano De~Cristofaro, Michael
  Sirivianos, Gianluca Stringhini, and Jeremy Blackburn.
\newblock Disinformation warfare: Understanding state-sponsored trolls on
  twitter and their influence on the web.
\newblock In {\em Companion proceedings of the 2019 world wide web conference},
  pages 218--226, 2019.

\end{thebibliography}

\clearpage

\appendix

\section{Pro-Ukrainian \vs Pro-Russian Hashtags}

\begin{tcolorbox}[colback=gray!5!white,colframe=gray!50!black]

\textbf{Pro-Ukraine:}

\#StandWithUkraine (333,999), \#StopPutin (130,300), \#StopRussia (96,761), \#StandWithUkriane (63,023), \#IStandWithUkraine (54,103), \#SafeAirLiftUkraine (49,168), \#StopPutinNow (34,985), \#StopRussianAggression (34,178), \#PutinIsAWarCriminal (29,502), \#SlavaUkraini (27,210), \#PutinHitler (26,153), \#PrayForUkraine (23,409), \#FckPutin (19,504), \#WeStandWithUkraine (18,388), \#StayWithUkraine (13,772), \#StandingWithUkraine (13,340), \#HelpUkraine (11,455), \#DefeatPutin (11,429), \#IStandWithUkriane (11,304), \#SaveUkraine (10,209), \#SanctionRussiaNow (9,067), \#RussiaGoHome (8,623), \#BanRussiaFromSwift (7,485), \#PrayingForUkraine (7,310), \#PutinWarCrimes (6,797), \#UnitedWithUkraine (6,697), \#FuckPutin (6,595), \#NoWarInUkraine (5,158), \#SolidarityWithUkraine (5,008), \#GloryToUkraine (4,377), \#BoycottRussia (4,292), \#StopRussianAgression (3,976), \#StandForUkraine (3,748), \#WeAreAllUkrainians (3,659), \#FckPtn (3,319), \#SendNatoToUkraine (3,275), \#PeaceForUkraine (2,421), \#StandWithUkraineNow (2,213)

\textbf{Pro-Russia:}

\#IStandWithPutin (134,393), \#IStandWithRussia (67,955), \#Hypocrisy (18,610), \#AbolishNato (14,404), \#RacistEU (12,276), \#StandWithRussia (9,451), \#DoubleStandards (7,883), \#ISupportRussia (5,929), \#NaziUkraine (4,467), \#StandWithPutin (3,228), \#SupportRussia (2,254)

\end{tcolorbox}

\end{document}